# A Physics-based and Data-driven Linear Three-Phase Power Flow Model for Distribution Power Systems

Yitong Liu, *Student Member, IEEE*, Zhengshuo Li, *Senior Member, IEEE*, Yu Zhou, *Student Member, IEEE*

*Abstract*—Distribution power systems (DPSs) are mostly unbalanced, and their loads may have notable static voltage characteristics (ZIP loads). Hence, despite abundant papers on linear single-phase power flow models, it is still necessary to study linear three-phase distribution power flow models. To this end, this paper proposes a physics-based and data-driven linear three-phase power flow model for DPSs. We first formulate how to amalgamate data-driven techniques into a physics-based power flow model to obtain our linear model. This amalgamation makes our linear model independent of the assumptions commonly used in the literature (e.g., nodal voltages are nearly 1.0 p.u.) and thus have a relatively high accuracy generally - even when those assumptions become invalid. We then reveal how to apply our model to the DPSs with ZIP loads. We also show that with the Huber penalty function employed, the adverse impact of bad data on our model's accuracy is significantly reduced, rendering our model robust against poor data quality. Case studies have demonstrated that our model generally has 2 to over 10-fold smaller average errors than other linear power flow models, enjoys a satisfying accuracy against bad data, and facilitates a faster solution to DPS analysis and optimization problems.

## I. INTRODUCTION

POWER flow analysis is fundamental to power system planning and operation, underlying contingency analysis, optimal dispatch and control, etc. [1]. However, the inherent nonlinearity of alternating current power flow (ACPF) equations makes it challenging to solve optimal power flow (OPF) problems for large-scale systems [2]. Indeed, it was proved in [3] that OPF with ACPF is NP-hard. Hence, practitioners often attempt to replace the ACPF with a linear power flow model that linearly and closely approximates the nonlinear relationships between nodal power and voltages.

A classical and widely-used linear power flow model is direct current power flow (DCPF). However, it is well known that the precondition for DCPF's accuracy mainly holds for transmission power systems (TPSs) [4] and will be invalid for distribution power systems (DPSs). Also, the classical DCPF is often deemed incompetent for the scenarios where systemwide reactive power and voltage issues are deeply concerned.

Recently, increasing penetration of distributed energy resources (DERs) has brought considerable attention to DPS operation problems and the coordination between DPSs and TPSs [5]. The infamous over-voltage issue and the fact that a three-phase DPS typically operates in an unbalanced condition with high branch resistance/reactance ratios render the classical DCPF unsuitable for DPSs' dispatch and control problems. To bridge this gap, researchers have proposed a bundle of linear distribution power flow models (including the modified DCPF models [6]) that can be roughly divided into the following three categories.

The linear models belonging to the first category are mainly derived from nodal power flow (or voltage) equations. There are two subcategories[1].

- The models in the first subcategory, exemplified by [6],[7], are constructed by (repeatedly) using the assumptions such as nearly identical nodal voltage magnitudes (i.e., $v_i \approx v_j \approx 1$), and/or nearly zero voltage phase angle differences across any branch (i.e., $\theta_{ij} \approx 0$). Ref. [1], though employing $\log(v)$ rather than $v$ as the state variables, also belongs to this subcategory. However, repeatedly using the above assumptions may result in approximation errors that reportedly can be relatively large in some cases.
- The models belonging to the second subcategory are constructed by using first-order Taylor expansions, e.g., [8]-[10]. Ref. [11] employs a tangent plane to approximate the manifold of power flow through smooth coordinate transformation of the state variables. Ref. [9] suggests choosing $v^\kappa$ as the state variables for linear single-phase power flow model which only caters to balanced three-phase DPSs, but a three-phase DPS is often unbalanced. Refs. [8] and [12] propose linear three-phase power flow models for DPSs, but the relationships between active or reactive power and voltage are not explicitly formulated.

The linear models in the second category are derived from branch power flow equations. There are also two subcategories.

This work was supported in part by the National Natural Science Foundation of China under Grant 52007105, Young Elite Scientists Sponsorship Program by CSEE under Grant JLB-2020-170, and Qilu Youth Scholar Program from Shandong University.
Y. Liu, Z. Li, and Y. Zhou are with Department of Electrical Engineering, Shandong University, Jinan 250061, China. Zhengshuo Li is the corresponding author (e-mail: zsli@sdu.edu.cn).

---

[1] Admittedly, this is a rough classification, simply to highlight the differences between our and others' works. Hence, the intersection of the subcategories might not be empty. In fact, [2] shows that some models in the first subcategory can be considered special instances of the first-order Taylor expansion models.

- The models in the first subcategory are derived from the classical DistFlow model [13] and its variants. A typical linear model neglects the quadratic-current related term that accounts for losses (e.g., [13]-[15]). An improved model is to linearize, rather than neglecting, the quadratic-current related term, as shown in [16] and [17].
- The second subcategory employs the first-order Taylor expansion of the branch power flow to derive the linear relationships between the branch power and the independent state variables. It is worth noting that the choices of these state variables are not unique, e.g., $v$ [18],[19], $v^2$ [20], or other variants (such as $v^2\theta$ in [21]). Ref. [2] provides theoretical results about the choices of the state variables for higher accuracy. However, when a linear model is employed in optimal dispatch and control problems, choosing proper state variables should also be subject to practical considerations, such as that the chosen state variables had better lead to linear and straightforward relationships between nodal power injections/branch power flow and nodal voltages.

The linear models in the third category employ *data-driven* techniques either instead of or in addition to the preceding *physics-based* linear models. This category has been attracting lots of attention owing to the fast-spreading of phasor measurement units (PMUs) and data acquisition systems [22],[23]. The historical operating data can be used to rebuild and/or linearize power system models. The construction of a linear power flow model in [24] is purely data-driven, requiring no *a priori* knowledge of grid models. Hence, its approximation errors depend entirely on the accuracy of the historical data. To partly reduce this dependence, [25] proposes a hybrid physical model-driven and data-driven approach. However, this approach employs $v^\kappa$ as state variables and might introduce nonlinearity in certain operational constraints in OPF or control problems where it is applied. Moreover, the model in [25] only caters to a balanced three-phase DPS, nor does it discuss how to mitigate the adverse impact of possible outliers in the data set, i.e., "bad data", a widely concerned issue in the real world [22].

This paper develops a physics-based and data-driven linear power flow model catering to both unbalanced and balanced three-phase DPSs. Compared with the prior studies, our linear model has the following features:

1) Unlike the studies only catering to balanced DPSs that are often dominated by unbalanced DPSs in the real world, our linear model caters to both unbalanced and balanced three-phase DPSs. Moreover, as data-driven techniques are amalgamated, our linear model is independent of the assumptions commonly used in the literature (e.g., nodal voltages are nearly 1.0 p.u.) and is generally more accurate. Besides, our model applies to the scenarios where constant impedance, constant current, and constant power (ZIP) load models must be considered to reflect the notable static voltage characteristics of loads.

2) In contrast to [2], [25], we consider the adverse impact of bad data and introduce a Huber penalty function to diminish this impact. Thus, our model is relatively accurate even against poor data quality - a widely concerned issue in data-driven related studies. Simulations in the sequel will show that our linear model is generally more accurate than several representative models in the literature, e.g., [2],[8],[13],[16].

3) Unlike those linear models employing nonlinear and complex functions of $v$, our model enjoys linear and straightforward relationships between nodal power and voltages, which facilitate itself readily being employed in DPS analysis and optimization problems. To show this, we provide an instance of evaluating the reactive power support range that a DSP can provide for its upstream TPS. The simulation shows that using our linear model facilitates a faster and relatively accurate solution to this reactive power support range evaluation problem.

The remainder of this paper will be arranged as follows. Section II presents our linear three-phase power flow model. Section III discusses some relevant issues such as handling bad data, applying our model to evaluate a DPS's reactive power support range, etc. Section IV presents simulation results to demonstrate our model's efficacy compared with other linear power flow models. Conclusions are drawn in Section V.

## II. LINEAR THREE-PHASE POWER FLOW MODEL

In the sequel, we first consider constant power loads and derive a complex linear three-phase DPS power flow model in terms of complex nodal power and voltages. Next, we derive real and linear relationships between three-phase active/reactive power and nodal voltages, which might be relatively easy to employ in DPS analysis and optimization problems. Then, we extend our model to involve ZIP loads.

Throughout the paper, symbols $\mathbb{C}$ and $\mathbb{R}$ denote the field of complex and real numbers, respectively; the superscripts T and $-1$ the transposition and inverse operator, respectively; $\mathbf{I}$ and $\mathbf{1}$ the identity matrix and all-one vector of proper dimensions, respectively; $\overline{\blacksquare}$ the conjugation of a complex vector or number; $diag(\blacksquare)$ the diagonal matrix generated by a vector; $|\blacksquare|$ the modulus of a complex number; $\|\blacksquare\|$ the sum of the moduli of the elements in a vector; $\widetilde{\blacksquare}$ the linear approximations of concerned variables; $./$ the element-wise division; $\jmath=\sqrt{-1}$.

### A. Complex Linear Model

Consider that a three-phase DPS with $n+1$ nodes numbered from 0 to $n$ is connected to the main grid at the slack node 0, i.e., the point of common coupling (PCC). The PCC nodal three-phase voltage $\mathbf{v}_0 = (v_0^a, v_0^b, v_0^c)^\mathrm{T} \in \mathbb{C}^3$ is known, and the current and power at this node are $\mathbf{i}_0 = (i_0^a, i_0^b, i_0^c)^\mathrm{T} \in \mathbb{C}^3$ and $\mathbf{s}_0 = (s_0^a, s_0^b, s_0^c)^\mathrm{T} \in \mathbb{C}^3$, respectively. The nodes 1 to $n$ are modeled as PQ-type nodes[2], whose complex nodal voltages, currents, and power are $\mathbf{v}_L = (\mathbf{v}_1^\mathrm{T}, \ldots, \mathbf{v}_N^\mathrm{T})^\mathrm{T} \in \mathbb{C}^{3n}$, $\mathbf{i}_L = (\mathbf{i}_1^\mathrm{T}, \ldots, \mathbf{i}_N^\mathrm{T})^\mathrm{T} \in \mathbb{C}^{3n}$, and $\mathbf{s}_L = (\mathbf{s}_1^\mathrm{T}, \ldots, \mathbf{s}_n^\mathrm{T}) \in \mathbb{C}^{3n}$, respectively. The current and power vectors of a node without loads or generation are set to zeros.

---

[2] Discussions about the validity of this PQ-type node model in the face of distributed generators can be referred to in [26]'s Footnote 1.

Notice that previous works, e.g., [27], have shown that one can first derive three-phase power flow equations that cater to wye-connected loads and then involve delta-connected loads readily based on the relationships in (1) where the superscripts *ab*, *bc*, and *ca* stand for line-to-line variables. Hence, in deriving our linear power flow model below, we assume that all the loads and generators are wye-connected. Delta-connected loads can be involved in the same way as shown in [27].

$$\mathbf{v}_{L,i}^{ab,bc,ca} = \begin{bmatrix} 1 & -1 & 0 \\ 0 & 1 & -1 \\ -1 & 0 & 1 \end{bmatrix} \mathbf{v}_{L,i}^{a,b,c}, \mathbf{i}_{L,i}^{a,b,c} = \begin{bmatrix} 1 & 0 & -1 \\ -1 & 1 & 0 \\ 0 & -1 & 1 \end{bmatrix} \mathbf{i}_{L,i}^{ab,bc,ca}. \quad (1)$$

For a three-phase DPS, Ohm's law indicates $\mathbf{i} = \mathbf{Y}\mathbf{v}$, where $\mathbf{Y} \in \mathbb{C}^{3(n+1) \times 3(n+1)}$, the admittance matrix pertinent to all the $n+1$ nodes, comprises a series of following 3×3 submatrices[3],

$$\mathbf{Y}_{ij} = \begin{bmatrix} y_{ij}^{aa} & y_{ij}^{ab} & y_{ij}^{ac} \\ y_{ij}^{ba} & y_{ij}^{bb} & y_{ij}^{bc} \\ y_{ij}^{ca} & y_{ij}^{cb} & y_{ij}^{cc} \end{bmatrix}, \ i \in \{0,1,...,n\}, j \in \{0,1,...,n\}.$$

It follows from $\mathbf{i} = \mathbf{Y}\mathbf{v}$ that

$$\begin{bmatrix} \mathbf{i}_0 \\ \mathbf{i}_L \end{bmatrix} = \begin{bmatrix} \mathbf{Y}_{00} & \mathbf{Y}_{0L} \\ \mathbf{Y}_{L0} & \mathbf{Y}_{LL} \end{bmatrix} \begin{bmatrix} \mathbf{v}_0 \\ \mathbf{v}_L \end{bmatrix}, \quad (2)$$

where the subscripts 0 and $L$ denote the PCC and the other $n$ nodes, respectively, and $\mathbf{Y}_{00} \in \mathbb{C}^{3 \times 3}$, $\mathbf{Y}_{L0} \in \mathbb{C}^{3n \times 3}$, $\mathbf{Y}_{0L} \in \mathbb{C}^{3 \times 3n}$, $\mathbf{Y}_{LL} \in \mathbb{C}^{3n \times 3n}$.

By substituting $\mathbf{s}_L = diag(\mathbf{v}_L)\bar{\mathbf{i}}_L$ into (2), we have

$$\mathbf{v}_L = \mathbf{w} + \mathbf{Y}_{LL}^{-1}(diag(\bar{\mathbf{v}}_L)^{-1}\bar{\mathbf{s}}_L), \quad (3)$$

where $\mathbf{w} = -\mathbf{Y}_{LL}^{-1}\mathbf{Y}_{L0}\mathbf{v}_0$ is a constant vector.

Equation (3) indicates that the nonlinear relationships between $\mathbf{v}_L$ and $\mathbf{s}_L$ are due to the existence of the term $diag(\bar{\mathbf{v}}_L)^{-1}\bar{\mathbf{s}}_L$. A natural scheme for approximating this term is to replace the unknown $diag(\bar{\mathbf{v}}_L)^{-1}$ by its guessed value $diag(\bar{\hat{\mathbf{v}}}_L)^{-1}$. Thus, (3) can be approximated as

$$\tilde{\mathbf{v}}_L = \mathbf{w} + \mathbf{Y}_{LL}^{-1}\left(diag(\bar{\hat{\mathbf{v}}}_L)^{-1}\bar{\mathbf{s}}_L\right). \quad (4)$$

Unless this guessed value happens to be the solution to (3), i.e., $\hat{\mathbf{v}}_L = \mathbf{v}_L$, which rarely occurs, $\tilde{\mathbf{v}}_L$ solved from (4) should inevitably differ from $\mathbf{v}_L$. If we compute the nodal currents using $\tilde{\mathbf{v}}_L$, namely $\tilde{\mathbf{i}}_L = \mathbf{Y}_{L0}\mathbf{v}_0 + \mathbf{Y}_{LL}\tilde{\mathbf{v}}_L$, then we have an approximate nodal power $\tilde{\mathbf{s}}_L = diag(\mathbf{v}_L)\overline{\tilde{\mathbf{i}}_L}$. The difference between this $\tilde{\mathbf{s}}_L$ and the true $\mathbf{s}_L$ can be computed as follows:

*Resulting Nodal Power Difference*

$$= \mathbf{s}_L - diag(\mathbf{v}_L)\overline{\left\{\mathbf{Y}_{L0}\mathbf{v}_0 + \mathbf{Y}_{LL}\left[\mathbf{w} + \mathbf{Y}_{LL}^{-1}\left(diag(\bar{\hat{\mathbf{v}}}_L)^{-1}\bar{\mathbf{s}}_L\right)\right]\right\}} \quad (5)$$

$$= [\mathbf{I} - diag(\mathbf{v}_L./\hat{\mathbf{v}}_L)]\mathbf{s}_L = diag(\mathbf{1} - \mathbf{v}_L./\hat{\mathbf{v}}_L)\mathbf{s}_L.$$

It is straightforward to see that the difference in (5) reflects how much the guessed $\hat{\mathbf{v}}_L$ is different from $\mathbf{v}_L$. Furthermore, to prevent the impact of $\mathbf{s}_L$, whose values may vary significantly case by case, on the magnitude of the difference in (5), one can further define the *normalized nodal power difference* $\varepsilon(\mathbf{v}_L, \hat{\mathbf{v}}_L)$ shown in (6) as a candidate *indicator* of the errors in the linear approximation in (4):

$$\varepsilon(\mathbf{v}_L, \hat{\mathbf{v}}_L) = diag(\mathbf{1} - \mathbf{v}_L./\hat{\mathbf{v}}_L), \quad (6)$$

which is a function of $\mathbf{v}_L$ and $\hat{\mathbf{v}}_L$.

---

[3] As for a single- or two-phase segment, its admittance matrix is 1×1 or 2×2. However, following the practice in [28], we can insert zero rows and columns for the missing phases to keep the admittance matrix still a 3×3 matrix.

Apparently, the smaller the moduli of the complex numbers in $\varepsilon$ are, the closer $\hat{\mathbf{v}}_L$ is to $\mathbf{v}_L$, indicating a more accurate linear approximation in (4). However, since the true value $\mathbf{v}_L$ cannot be precisely known before solving (3), it is typically hard to select a "well-guessed" $\hat{\mathbf{v}}_L$ that can generally reduce the modulus of the element in $\varepsilon(\mathbf{v}_L, \hat{\mathbf{v}}_L)$ for all possible $\mathbf{v}_L$. This is reportedly a major reason why some physics-based linear power flow models yield unneglectable errors in certain cases.

Thanks to the spreading of data acquisition systems in DPSs, it becomes possible to learn a well-guessed $\hat{\mathbf{v}}_L$ from historical data: one can exploit data-driven techniques to reduce the above physics-based linear model's errors. The idea of this amalgamation is briefly stated below.

Suppose one has a series of independent historical data from which one can select ℵ typical operational states in terms of nodal voltages. Let $\hat{\mathbf{v}}_L$ be an element-wise linear combination of these historical voltages pertinent to the ℵ operational states. Then, optimize the associated combination coefficients using the remaining historical data (i.e., the training data), e.g., through a least-squares method, to generate a "well-guessed" $\hat{\mathbf{v}}_L$. To simplify the following formulae, we take ℵ = 2 as an example throughout this paper.

When ℵ = 2, one could choose the first typical operational state representing that this DPS operates under very light loading conditions and the other state very heavy loading conditions. Let $\hat{\mathbf{v}}_u$ and $\hat{\mathbf{v}}_l$ denote the voltage vectors pertinent to these two states. Then, the term $diag(\bar{\hat{\mathbf{v}}}_L)^{-1}$ in (4) can be represented by an element-wise linear combination of the counterparts pertinent to $\hat{\mathbf{v}}_u$ and $\hat{\mathbf{v}}_l$. Thus, we have a new linear three-phase power flow model below:

$$\tilde{\mathbf{v}}_L = \mathbf{w} + \mathbf{Y}_{LL}^{-1}\left(\boldsymbol{\mu} diag(\bar{\hat{\mathbf{v}}}_u)^{-1} + (\mathbf{I}-\boldsymbol{\mu})diag(\bar{\hat{\mathbf{v}}}_l)^{-1}\right)\bar{\mathbf{s}}_L, \quad (7)$$

where $\boldsymbol{\mu}$ is a real diagonal matrix with $3n$ diagonal elements that are the real combination coefficients to be learned from the training data.

In order to seek an optimal $\boldsymbol{\mu}$, define the following normalized nodal power difference related to the model (7):

$$\varepsilon(\mathbf{v}_L, \boldsymbol{\mu}) = \mathbf{I} - [\boldsymbol{\mu} diag(\mathbf{v}_L./\hat{\mathbf{v}}_u) + (\mathbf{I}-\boldsymbol{\mu})diag(\mathbf{v}_L./\hat{\mathbf{v}}_l)], \quad (8)$$

which is a function of $\mathbf{v}_L$ and $\boldsymbol{\mu}$.

Since we aim to find a $\boldsymbol{\mu}$ that can reduce the moduli of the elements in $\varepsilon(\mathbf{v}_L, \boldsymbol{\mu})$ for all possible $\mathbf{v}_L$, a natural scheme is to resort to a least-squares method that yields an optimal $\boldsymbol{\mu}_{\text{opt}}$ by solving the unconstrained optimization (9) that is constructed by using the training data:

$$\boldsymbol{\mu}_{\text{opt}} = \underset{\boldsymbol{\mu}}{\operatorname{argmin}} \sum_k \|\varepsilon(\hat{\mathbf{v}}_k, \boldsymbol{\mu})\|, \quad (9)$$

where $k$ indicates the $k$-th training data. By substituting $\boldsymbol{\mu}_{\text{opt}}$ into (7), we finally obtain our physics-based and data-driven linear three-phase power flow model that shows linear relationships between complex nodal voltages and power injections.

### B. Real Linear Model

The linear model (7) is formulated in terms of complex nodal power and voltages. However, it can be readily recast into a real linear three-phase model to reveal the linear relationships between nodal active/reactive power and nodal voltages, as is shown below.

Continue to use the symbols in Section II.A. Besides, let $\mathbf{v}_i = [v_i^a, v_i^b, v_i^c]^T$ and $\mathbf{s}_i = \mathbf{p}_i + j\mathbf{q}_i = [p_i^a, p_i^b, p_i^c]^T + j[q_i^a, q_i^b, q_i^c]^T$ be the three-phase voltage and power at node $i$ ($i \neq 0$), respectively; the complex admittance matrices $\mathbf{Y}_{L0} = \mathbf{G}_{L0} + j\mathbf{B}_{L0}$, $\mathbf{Y}_{LL} = \mathbf{G}_{LL} + j\mathbf{B}_{LL}$.

Like what we did in Section II.A, one can approximate node $i$'s three-phase active and reactive power using this node's approximate current $\tilde{\mathbf{i}}_{L,i} = \tilde{\mathbf{i}}_{re,i} + j\tilde{\mathbf{i}}_{im,i}$ as follows,

$$\mathbf{p}_i + j\mathbf{q}_i = diag(\mathbf{v}_{re,i} + j\mathbf{v}_{im,i})(\tilde{\mathbf{i}}_{re,i} - j\tilde{\mathbf{i}}_{im,i}) \quad (10)$$

where the subscripts $re$ and $im$ represent the real and imaginary parts of the complex three-phase $\mathbf{v}_i$ and $\tilde{\mathbf{i}}_{L,i}$, respectively. All $\tilde{\mathbf{i}}_{L,i}$ for nodes 1 to $n$ comprise $\tilde{\mathbf{i}}_L = (\mathbf{G}_{L0,i} + j\mathbf{B}_{L0,i})\mathbf{v}_0 + \sum_{j=1}^n (\mathbf{G}_{LL,ij} + j\mathbf{B}_{LL,ij})\hat{\mathbf{v}}_j$, an approximation to $\mathbf{i}_L$ by replacing the unknown $\mathbf{v}_L$ with a guessed $\hat{\mathbf{v}}$ in (2), where $\hat{\mathbf{v}}_j$ is node $j$'s approximate three-phase voltage that comprises $\hat{\mathbf{v}}$, and $\mathbf{G}_{L0,i}, \mathbf{B}_{L0,i}, \mathbf{G}_{LL,ij}$, and $\mathbf{B}_{LL,ij}$ are 3×3 matrices pertinent to node $i$ in the format of

$$\mathbf{M}_x = \begin{bmatrix} m_x^{aa} & m_x^{ab} & m_x^{ac} \\ m_x^{ba} & m_x^{bb} & m_x^{bc} \\ m_x^{ca} & m_x^{cb} & m_x^{cc} \end{bmatrix},$$

where $\mathbf{M}$ denotes $\mathbf{G}$ or $\mathbf{B}$; $m$ denotes the conductance $g$ or susceptance $b$; and the subscript $x$ denotes the composite symbols $L0, i$ or $LL, ij$.

In order to achieve a generally high approximation accuracy, following the above idea of amalgamating data-driven techniques into a physics-based model, the guessed $\hat{\mathbf{v}}$ should be subject to $diag(\overline{\hat{\mathbf{v}}})^{-1} = \boldsymbol{\mu}_{opt} diag(\overline{\hat{\mathbf{v}}}_u)^{-1} + (\mathbf{I} - \boldsymbol{\mu}_{opt}) diag(\overline{\hat{\mathbf{v}}}_l)^{-1}$ where the real matrix $\boldsymbol{\mu}_{opt}$ can be solved from (9), just as what we did in Section II.A.

Moreover, given (10), it is easy to compute the active and reactive power for phase $\varphi$ of node $i$ as follows,

$$p_i^\varphi = \tilde{i}_{re,i}^\varphi v_{re,i}^\varphi + \tilde{i}_{im,i}^\varphi v_{im,i}^\varphi, \quad q_i^\varphi = \tilde{i}_{re,i}^\varphi v_{im,i}^\varphi - \tilde{i}_{im,i}^\varphi v_{re,i}^\varphi, \quad (11)$$

which indicate that $p_i^\varphi$ and $q_i^\varphi$ are linear in nodal voltage variables $v_{re,i}^\varphi$ and $v_{re,i}^\varphi$ where $\tilde{i}_{re,i}^\varphi, \tilde{i}_{im,i}^\varphi$ are set to

$$\tilde{i}_{re,i}^\varphi = \sum_{\varphi'} \left[ (g_{L0,i}^{\varphi\varphi'} v_{re,0}^{\varphi'} - b_{L0,i}^{\varphi\varphi'} v_{im,0}^{\varphi'}) + \sum_{j=1}^n (g_{LL,ij}^{\varphi\varphi'} \hat{v}_{re,j}^{\varphi'} - b_{L0,ij}^{\varphi\varphi'} \hat{v}_{im,j}^{\varphi'}) \right]$$

$$\tilde{i}_{im,i}^\varphi = \sum_{\varphi'} \left[ (g_{L0,i}^{\varphi\varphi'} v_{im,0}^{\varphi'} + b_{L0,i}^{\varphi\varphi'} v_{re,0}^{\varphi'}) + \sum_{j=1}^n g_{LL,ij}^{\varphi\varphi'} \hat{v}_{im,j}^{\varphi'} + b_{L0,ij}^{\varphi\varphi'} \hat{v}_{re,j}^{\varphi'} \right]$$

and the superscript $\varphi'$ is arranged in the order of a, b, and c.

Thus, we have obtained the desired real linear three-phase model (11).

### C. To Involve ZIP Loads

The above linear power flow model can be extended to involve ZIP loads. As explained earlier, we assume that node $i$ ($i \neq 0$) is connected to a wye-connected ZIP load that is commonly formulated as follows,

$$p_i^\varphi = \lambda_{p,i}^\varphi p_{nom,i}^\varphi \left[ a_{p,i}^\varphi \left( \frac{|v_i^\varphi|}{|v_{nom,i}^\varphi|} \right)^2 + b_{p,i}^\varphi \frac{|v_i^\varphi|}{|v_{nom,i}^\varphi|} + c_{p,i}^\varphi \right],$$

$$q_i^\varphi = \lambda_{q,i}^\varphi q_{nom,i}^\varphi \left[ a_{q,i}^\varphi \left( \frac{|v_i^\varphi|}{|v_{nom,i}^\varphi|} \right)^2 + b_{q,i}^\varphi \frac{|v_i^\varphi|}{|v_{nom,i}^\varphi|} + c_{q,i}^\varphi \right], \quad (12)$$

where $\varphi$ denotes phase a, b, or c; $\lambda_{p,i}^\varphi$ and $\lambda_{q,i}^\varphi$ the *nominal active and reactive power change rate* for phase $\varphi$ of node $i$ with respect to the nominal active and reactive power $p_{nom,i}^\varphi$ and $q_{nom,i}^\varphi$, respectively; $v_{nom,i}^\varphi$ the nominal nodal voltage; $v_i^\varphi$ the actual voltage; $a_{p,i}^\varphi, b_{p,i}^\varphi$, and $c_{p,i}^\varphi$ are the real coefficients of each component in the ZIP loads, and their sum is 1; the same are $a_{q,i}^\varphi, b_{q,i}^\varphi$, and $c_{q,i}^\varphi$.

To simplify the following derivation, we also assume that nodal active and reactive power alter with a *fixed power factor* as well as the same impacts of the static voltage characteristics on the active and reactive power for the same phase of the same node, which are formally formulated below:

*Assumption 1:* $a_{p,i}^\varphi = a_{q,i}^\varphi = a_i^\varphi$, $b_{p,i}^\varphi = b_{q,i}^\varphi = b_i^\varphi$, $c_{p,i}^\varphi = c_{q,i}^\varphi = c_i^\varphi$, $\lambda_{p,i}^\varphi = \lambda_{q,i}^\varphi = \lambda_i^\varphi$ for the same $\varphi$ and the same $i$, and they are all real numbers.

Then, the complex power for phase $\varphi$ of node $i$ is

$$\begin{aligned} s_i^\varphi &= p_i^\varphi + jq_i^\varphi \\ &= \lambda_i^\varphi (p_{nom,i}^\varphi + jq_{nom,i}^\varphi) \left[ a_i^\varphi \left( \frac{|v_i^\varphi|}{|v_{nom,i}^\varphi|} \right)^2 + b_i^\varphi \frac{|v_i^\varphi|}{|v_{nom,i}^\varphi|} + c_i^\varphi \right] \quad (13) \\ &= \lambda_i^\varphi s_{nom,i}^\varphi A(|v_i^\varphi|, |v_{nom,i}^\varphi|), \end{aligned}$$

which is a function of $\lambda_i^\varphi$ and $v_i^\varphi$. $A(|v_i^\varphi|, |v_{nom,i}^\varphi|)$ defined in (13) is a real function and $s_{nom,i}^\varphi$ the nominal nodal complex power. Now that $s_i^\varphi$ is contingent on $\lambda_i^\varphi$ and $v_i^\varphi$, we are turning to derive the linear relationships between $v_i^\varphi$ and $\lambda_i^\varphi$.

Denote the *systemwide nominal nodal power change rate* by $\boldsymbol{\lambda} = [\lambda_i^\varphi]_{i=1...N, \varphi=a,b,c} \in \mathbb{R}^{3n}$. Substituting (13) into (4) with a guessed term $\hat{\mathbf{v}}_L$, one obtains the following linear three-phase power flow model:

$$\tilde{\mathbf{v}}_L = \mathbf{w} + \mathbf{Y}_{LL}^{-1} \left( diag(\mathbf{B}(|\hat{\mathbf{v}}_L|, |\mathbf{v}_{nom}|)) diag(\bar{\mathbf{s}}_{L0}) \right) \boldsymbol{\lambda}, \quad (14)$$

where $\mathbf{v}_{nom}$ is the vector of $v_{nom,i}^\varphi$, $\varphi = a, b, c$, $i = 1, ..., n$; $diag(\mathbf{B}(|\hat{\mathbf{v}}_L|, |\mathbf{v}_{nom}|)) = diag(\overline{\hat{\mathbf{v}}}_L)^{-1} diag(A(|\hat{\mathbf{v}}_L|, |\mathbf{v}_{nom}|))$, where $A(|\hat{\mathbf{v}}_L|, |\mathbf{v}_{nom}|)$ is the vector consisting of $A(|v_i^\varphi|, |v_{nom,i}^\varphi|)$, $\varphi = a, b, c$, $i = 1, ..., n$.

It is easy to verify that the formula of the nodal power difference pertinent to (14) is similar to (5), so one can also amalgamate the data-driven techniques here to improve the model's accuracy. After choosing proper historical points $\hat{\mathbf{v}}_u$ and $\hat{\mathbf{v}}_l$ to represent the guessed $\hat{\mathbf{v}}_L$, we have

$$\tilde{\mathbf{v}}_L = \mathbf{w} + \mathbf{Y}_{LL}^{-1} [\boldsymbol{\mu} diag(\mathbf{B}(|\hat{\mathbf{v}}_u|, |\mathbf{v}_{nom}|)) \\ + (\mathbf{I} - \boldsymbol{\mu}) diag(\mathbf{B}(|\hat{\mathbf{v}}_l|, |\mathbf{v}_{nom}|))] diag(\bar{\mathbf{s}}_{L0}) \boldsymbol{\lambda} \quad (15)$$

The resultant nodal power difference is $\mathbf{s}_L - \tilde{\mathbf{s}}_L = (\mathbf{I} - \mathbf{K})\mathbf{s}_L$ where $\mathbf{K}$ is defined in (16).

$$\begin{aligned} \mathbf{K} &= \boldsymbol{\mu} diag(\mathbf{v}_L./\hat{\mathbf{v}}_u) diag(A(|\hat{\mathbf{v}}_u|, |\mathbf{v}_{nom}|)./A(|\mathbf{v}_L|, |\mathbf{v}_{nom}|)) \\ &+ (\mathbf{I} - \boldsymbol{\mu}) diag(\mathbf{v}_L./\hat{\mathbf{v}}_l) diag(A(|\hat{\mathbf{v}}_u|, |\mathbf{v}_{nom}|)./A(|\mathbf{v}_L|, |\mathbf{v}_{nom}|)). \end{aligned} \quad (16)$$

Similarly, define the normalized nodal power difference

$$\varepsilon(\mathbf{v}_L, \boldsymbol{\mu}) = (\mathbf{I} - \mathbf{K}). \quad (17)$$

Then, one can use the training data to construct a least-squares problem like (9) to solve an optimal $\boldsymbol{\mu}_{opt}$ to minimize the moduli of the elements in $\varepsilon$ in (17). By substituting this $\boldsymbol{\mu}_{opt}$ into (15), one then obtains a linear three-phase power flow model that caters to ZIP loads.

Notice that when $a_i^\varphi = b_i^\varphi = 0$, $c_i^\varphi = 1$ (i.e., constant power loads) throughout the DPS, we have $A(|v_i^\varphi|, |v_{nom,i}^\varphi|) = 1$ and $s_i^\varphi = \lambda_i^\varphi s_{nom,i}^\varphi$ so (15) is reduced to (7), as one would expect.

It is worth mentioning that when Assumption 1 is removed, one can still follow the above derivation to obtain a linear power flow model like (15) except that $A$ is now a complex function and $\boldsymbol{\lambda} \in \mathbb{R}^{6n}$ comprises $\lambda_{p,i}^\varphi$ and $\lambda_{q,i}^\varphi$.

## III. Discussions

### A. Dealing with Bad Data

The quality of historical data may affect the accuracy of our linear power flow model. In fact, despite field measurements' accuracy being improved recently, bad data might still possibly exist in a DPS's history database.

To diminish the adverse impact of bad data on evaluating $\boldsymbol{\mu}_{opt}$ and the resulting accuracy of our model, one can employ the Huber penalty function (18) in the least-squares problem (9) instead of the previous vanilla $\|\varepsilon(\hat{\mathbf{v}}_k, \boldsymbol{\mu})\|$.

$$\phi_{hub}(\|\varepsilon(\hat{\mathbf{v}}_k, \boldsymbol{\mu})\|) = \begin{cases} \|\varepsilon(\hat{\mathbf{v}}_k, \boldsymbol{\mu})\|^2 & \|\varepsilon(\hat{\mathbf{v}}_k, \boldsymbol{\mu})\| \leq \delta \\ \delta(2\|\varepsilon(\hat{\mathbf{v}}_k, \boldsymbol{\mu})\| - \delta) & \|\varepsilon(\hat{\mathbf{v}}_k, \boldsymbol{\mu})\| \geq \delta \end{cases}, \quad (18)$$

where $\varepsilon(\hat{\mathbf{v}}_k, \boldsymbol{\mu})$ denotes the complex error vector pertinent to the $k$-th training data; $\delta$ is a given threshold.

On the one hand, it is easy to see that when the residuals, i.e., $\|\varepsilon(\hat{\mathbf{v}}_k, \boldsymbol{\mu})\|$, are smaller than $\delta$, the Huber penalty function (18) is reduced to the original objective function in (9). On the other hand, the Huber penalty function penalizes those residuals larger than $\delta$ which are typically caused by outliers. The principle for setting $\delta$ can be found in [29]. In theory, the Huber penalty function has a desirable effect in dealing with the bad data [30] and a least-squares problem with the Huber penalty function defined in (18) can be readily recast as an equivalent convex problem which is easy to solve [31].

### B. Leveraging Forgetting Factors

As shown above, the weights pertinent to each training data are set equally by default. Nevertheless, the DPS operating state varies with time physically, so it would be more reasonable to consider that the historical data with different timestamps have different relevance to, and thus different impact on, the contemporary operating point [32]. To reflect this difference on evaluating $\boldsymbol{\mu}_{opt}$, one can adopt the approach of forgetting factors that are widely used in signal estimation to reflect the distance of the sampling time from the contemporary moment [33]. A common practice is to set larger weights for the historical data closer to the contemporary operating point. Because of the space limit, a detailed weight setting scheme is omitted here; interested readers can refer to [34].

### C. Application to Reactive Power Support Range Evaluation

Below we will briefly discuss how to employ our linear model to facilitate DPS analysis and optimization problems. We take an example of evaluating the reactive power support range that a DSP can provide for its upstream TPS.

Evaluating this reactive power support range becomes relevant as increasing penetration of DERs causes a shortage of reactive power sources in TPSs [35]. Ref. [35] presents an accurate but complicated method to evaluate this range in the presence of uncertainty accurately. If the uncertainty is neglectable, then our linear power flow model could be employed to make a faster evaluation.

We follow the convention in [35] that a high-voltage DPS with the PCC node numbered one is deemed balanced and modeled as a single-phase system. Our objective is to capture the reactive power range at the PCC node, i.e., $[\underline{q}_1, \overline{q}_1]$, while meeting all the commonly operational constraints regarding this DPS, as listed in (19),

$$\begin{aligned} v_{j,min}^2 &\leq |v_j|^2 \leq v_{j,max}^2, \quad \forall j \in \mathcal{N}\setminus\{1\}, \\ p_{j,min} &\leq p_j \leq p_{j,max}, \quad \forall j \in \mathcal{N}\setminus\{1\}, \\ q_{j,min} &\leq q_j \leq q_{j,max}, \quad \forall j \in \mathcal{N}\setminus\{1\}, \\ 0 &\leq |i_{ij}|^2 \leq i_{ij,max}^2, \quad \forall ij \in \mathcal{L}, \end{aligned} \quad (19)$$

where $\mathcal{L}$ and $\mathcal{N}$ denote the sets of the branches and nodes, respectively; the pairs of $v_{j,max}$ and $v_{j,min}$, $p_{j,max}$ and $p_{j,min}$, $q_{j,max}$ and $q_{j,min}$ are the upper and lower limits of the nodal voltage magnitude, generator's active power and reactive power, respectively; $i_{ij,max}$ is the upper limit of branch $ij$'s current.

The voltage magnitude constraints in (19) are quadratic in nodal voltages' real and imaginary parts. We linearize these quadratic constraints as shown in (20), where $\hat{\mathbf{v}}$, as discussed earlier, is subject to $diag(\overline{\hat{\mathbf{v}}})^{-1} = \boldsymbol{\mu}_{opt} diag(\overline{\hat{\mathbf{v}}}_u)^{-1} + (\mathbf{I} - \boldsymbol{\mu}_{opt})diag(\overline{\hat{\mathbf{v}}}_l)^{-1}$ and $\boldsymbol{\mu}_{opt}$ is solved from (9).

$$|v_j|^2 = \mathcal{R}(v_j) \cdot \hat{\mathbf{v}}_{re,j} + \mathcal{J}(v_j) \cdot \hat{\mathbf{v}}_{im,j}, \quad (20)$$

where $\mathcal{R}$ and $\mathcal{J}$ represent the operator of taking the real part and imaginary part of a complex number.

Similar linearization is applied to branches' current magnitude constraints in (19), where $|i_{ij}|^2 = [\mathcal{R}(v_i - v_j)(\hat{\mathbf{v}}_{re,i} - \hat{\mathbf{v}}_{re,j}) + \mathcal{J}(v_i - v_j) \cdot (\hat{\mathbf{v}}_{im,i} - \hat{\mathbf{v}}_{im,j})](g_{ij}^2 + b_{ij}^2)$.

Now that the nonlinear constraints in (19) have been linearized and that (11) has linearly approximated the relationships between the nodal active/reactive power and voltages, the DPS can be modeled as a linear system $\mathbf{A}_{sys}\mathbf{x}_{sys} \leq \mathbf{b}_{sys}$, where $\mathbf{x}_{sys}$ represents all the unknown variables in the DPS. Hence, solving the DPS's reactive power range $[\underline{q}_1, \overline{q}_1]$ is, mathematically, to project this linear system's feasible region onto the one-dimensional space of the PCC nodal reactive power, which is readily solvable, e.g., through the Fourier–Motzkin elimination [36].

### D. The Necessity of Employing a Physics-based Model

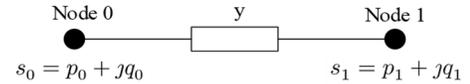

Fig. 1. A two-bus system.

This part will illustrate the necessity of employing a physics-based model on top of a purely data-driven model. Consider the balanced two-node system in Fig. 1. Let the voltages of Nodes 0 and 1 be 1.0 p.u. and $v_{1,*}$, respectively.

First, consider a purely data-driven model (21) where one has chosen proper historical data $\hat{v}_{1,u}$ and $\hat{v}_{1,l}$, and solved $\mu_{opt}$ from a least-squares problem, i.e., $\mu_{opt} = \arg\min_\mu \sum_k \|\tilde{v}_{1,d} - v_{1,k}\|$, where $v_{1,k}$ represents the $k$-th training data.

$$\tilde{v}_{1,d} = \mu\hat{v}_{1,u} + (1-\mu)\hat{v}_{1,l} \tag{21}$$

Next, by substituting $\tilde{v}_{1,d}$ into (3), one obtains $\tilde{v}_{1,h}$ in (22.a), which is a new approximation to $v_*$ by amalgamating the physics-based model. The true value of $v_{1,*}$ is given in (22.b).

$$\tilde{v}_{1,h} = w + Y_{LL}^{-1}\bar{s}_L(\bar{\tilde{v}}_{1,d}^{-1}) \tag{22.a}$$

$$v_{1,*} = w + Y_{LL}^{-1}\bar{s}_L\bar{v}_{1,*}^{-1} \tag{22.b}$$

From Fig. 1, we have $Y_{LL} = y$, $s_L = s_1$, so

$$\left|\tilde{v}_{1,h} - v_{1,*}\right| = \left|\frac{1}{y}\bar{s}_1(\bar{\tilde{v}}_{1,d}^{-1} - \bar{v}_{1,*}^{-1})\right| = \left|\frac{1}{\bar{y}}s_1(\tilde{v}_{1,d}^{-1} - v_{1,*}^{-1})\right|$$
$$= \left|\frac{1}{\bar{y}}s_1\right|\left|\tilde{v}_{1,d}^{-1} - v_{1,*}^{-1}\right| = \frac{\left|\frac{1}{\bar{y}}s_1\right|}{|\tilde{v}_{1,d}v_{1,*}|}|\tilde{v}_{1,d} - v_{1,*}|. \tag{23}$$

Since $|v_*|$ normally stay within 0.95 to 1.05 p.u., $|\bar{v}_{1,*} - 1|$ is far less than $|\tilde{v}_{1,d}|$, we have

$$\left|\frac{1}{\bar{y}}s_1\right| = \left|\frac{1}{\bar{y}}(v_*(-\bar{y} + \bar{y}\bar{v}_*))\right|$$
$$= |v_*(\bar{v}_* - 1)| \ll |v_*\tilde{v}_{1,d}|, \tag{24}$$

and (25) as well, which implies that amalgamating the physics-based model significantly reduces the errors in a purely data-driven model. That is why we amalgamate a physics-based model and data-driven techniques to construct our linear model.

$$|\tilde{v}_{1,h} - v_{1,*}| \ll |\tilde{v}_{1,d} - v_{1,*}| \tag{25}$$

## IV. CASE STUDIES

First, we compared our model with several representative linear models reported in [2],[8],[13],[16], on both single-phase (i.e., balanced) and unbalanced three-phase DPSs with both constant power and ZIP loads. Next, we considered bad data and showed the effect of employing the Huber penalty function in diminishing the impact of bad data. Finally, we tested how employing our model could facilitate solving the DPS reactive power support range evaluation problem.

Except for one test system from [35], all the other test systems are from MATPOWER and OpenDSS, whose power flow engines were used to generate the historical data by randomly changing the loads to construct our model and to test the accuracy of different linear models. Throughout this section, the linearization errors are measured in terms of the average relative errors of an array of our concerned variables, e.g., nodal voltages. The true values of the concerned variables in balanced DPSs were solved via MATPOWER and those in the unbalanced DPSs were solved via OpenDSS.

### A. Tests on Balanced DPSs with Constant Power Loads

#### 1) Compared with Physics-based Models

We compared our proposed linear model with the lossless DistFlow model [13] and the lossy Distflow model [16]. Three systems, 22-bus system, 85-bus system, and 141-bus system from MATPOWER, were used with the training set size being 100, 300, and 600, respectively. The test set size is set to 900. This training and test sets' set-up was referred to [25], where the test set size is larger than that of the training set. We think one reason for this set-up is to make a "worst-scenario" test to see whether a linear model has a generally high accuracy without a relatively large training set.

Table I shows three models' average relative errors pertinent to the nodal voltages, which indicates that our model is the most accurate as its average relative error was 10- to 100-fold smaller. The box plots in Fig. 2 present the relative error distributions, where our model's largest linearization errors are often below the smallest linearization errors of the other two models, firmly demonstrating our model's superior accuracy.

TABLE I
AVERAGE RELATIVE ERRORS OF THREE LINEAR MODELS

|  | Lossless Distflow [13] | Lossy Distflow [16] | Our model |
|---|---|---|---|
| 22-bus system | 2.4×10⁻³ | 2.2×10⁻³ | 5.58×10⁻⁵ |
| 85-bus system | 7.3×10⁻³ | 5.7×10⁻³ | 8.37×10⁻⁴ |
| 141-bus system | 9.1×10⁻³ | 9.0×10⁻³ | 1.89×10⁻⁴ |

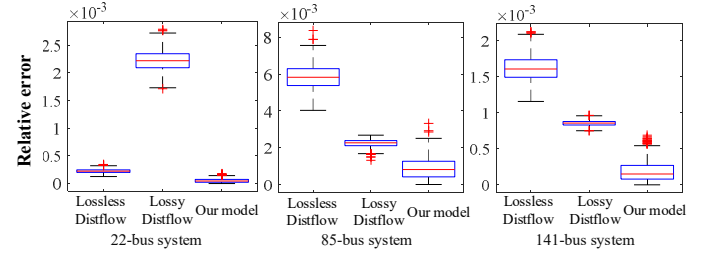

Fig. 2. Relative error distributions pertinent to the nodal voltages.

#### 2) Compared with Physics-based and Data-driven Model [2]

We further compared our model with the single-phase model [2], where the authors proposed to employ historical data to select a proper value of $\kappa$ in the term $v^\kappa$ to minimize the errors in the linearized active power $\tilde{p}_{ij}$ and reactive power $\tilde{q}_{ij}$ through branch $ij$, as shown in (26), where $\theta_{ij}$ is the voltage phase angle difference across this branch.

$$\tilde{p}_{ij} = g_{ij}\frac{(v_i^\kappa - v_j^\kappa)}{\kappa} - b_{ij}\theta_{ij}, \tilde{q}_{ij} = -b_{ij}\frac{(v_i^\kappa - v_j^\kappa)}{\kappa} - g_{ij}\theta_{ij}. \tag{26}$$

Different from [2], however, our model is to minimize the errors in nodal voltages. In this sense, a fair comparison of the two models' accuracy might be comparing the errors in the branch power flow yielded by the model [2] and those in nodal voltages by our model. Because the per-unit values of certain branches' power flow can be close to zero (e.g., a light loaded line) while the per-unit nodal voltages are always within 0.95-1.05 p.u., we adopted the *normalized relative errors* defined in (27) for $k$-th test data in this comparison.

$$err_{p_{ij},nor}^k = \frac{\epsilon_{p_{ij}}^k - \min_{k'\in\mathcal{K}}(\epsilon_{p_{ij}}^{k'})}{\max_{k'\in\mathcal{K}}(\epsilon_{p_{ij}}^{k'}) - \min_{k'\in\mathcal{K}}(\epsilon_{p_{ij}}^{k'})},$$
$$err_{q_{ij},nor}^k = \frac{\epsilon_{q_{ij}}^k - \min_{k'\in\mathcal{K}}(\epsilon_{q_{ij}}^{k'})}{\max_{k'\in\mathcal{K}}(\epsilon_{q_{ij}}^{k'}) - \min_{k'\in\mathcal{K}}(\epsilon_{q_{ij}}^{k'})}, \tag{27}$$
$$err_{v,nor}^k = \frac{\epsilon_v^k - \min_{k'\in\mathcal{K}}(\epsilon_v^{k'})}{\max_{k'\in\mathcal{K}}(\epsilon_v^{k'}) - \min_{k'\in\mathcal{K}}(\epsilon_v^{k'})},$$

where $\epsilon_{p_{ij}}^k = \sum_{ij\in\mathcal{L}}|\tilde{p}_{ij} - p_{ij,actual}|$, $\epsilon_{q_{ij}}^k = \sum_{ij\in\mathcal{L}}|\tilde{q}_{ij} - q_{ij,actual}|$, $\epsilon_v^k = \sum_{i\in\mathcal{N}}|\tilde{v}_i - v_{i,actual}|$, and $\mathcal{K}$ denotes the test set. Thus, $err_{p_{ij},nor}^k$, $err_{q_{ij},nor}^k$ in (27) were computed using the results from the model in [2], and $err_{v,nor}^k$ was computed using the results from our model.

Table II presents the average normalized relative value: $\sum_k err_{p_{ij},nor}^k/\text{K}$, $\sum_k err_{q_{ij},nor}^k/\text{K}$, and $\sum_k err_{v,nor}^k/\text{K}$, where K denotes the size of the test set. This table shows that our linear model's errors are generally half of the model [2] for single-phase DPSs.

Moreover, as discussed previously, we can further reduce our model's errors by setting proper forgetting factors. For example, if the forgetting factors were chosen from the range

0.7 to 1.0 based on how far in the timeline the training data are from the contemporary moment [33], our model's errors were further half smaller. A detailed weight setting scheme can be referred to [34].

TABLE II
COMPARISON WITH MODEL [2]

|  | $\frac{\sum_k err^k_{p_{ij},nor}}{K}$ | $\frac{\sum_k err^k_{q_{ij},nor}}{K}$ | $\frac{\sum_k err^k_{v,nor}}{K}$ |
|---|---|---|---|
| 22-bus system | 0.4488 | 0.4488 | 0.3279 |
| 85-bus system | 0.4235 | 0.4225 | 0.2485 |

### B. Tests on Unbalanced DPSs with ZIP Loads

Unlike the models in [13], [16], and [2] that either cater to balanced systems or neglect the static voltage characteristics of loads, our model caters to unbalanced DPSs with ZIP loads. To reveal our model's accuracy in this scenario, we tested it on three systems, IEEE 13-bus, 37-bus, and 123-bus systems, where the originally distributed load is divided into two equivalent lumped loads at the ends of a branch.

The average relative errors are listed in Table III, and the error distribution pertinent to the test sets is presented in Fig. 3. These minor errors confirm that our linear model has a satisfying accuracy for unbalanced DPSs even when loads have notable static voltage characteristics, which is notable merit that some linear single-phase power flow models might not naturally have.

Besides, we compared our model with the three-phase linear model in [8] on the 123-bus system. The average error of that model is $5.2 \times 10^{-2}$, about 10-fold larger than ours. In an initial investigation, we found several one- or two-phase segments in this 123-bus system, which may be why the model in [8] had relatively larger errors in this test. This test, in turn, suggests that our method's accuracy is robust against the issue of missing phases, which is relatively commonplace in real-world DPSs.

TABLE III
OUR MODEL'S AVERAGE RELATIVE ERROR FOR THREE THREE-PHASE DPSS

|  | 13 bus system | 37 bus system | 123 bus system |
|---|---|---|---|
| Errors | $1.35 \times 10^{-3}$ | $5.03 \times 10^{-3}$ | $6.56 \times 10^{-3}$ |

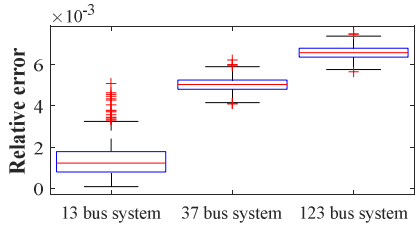

Fig. 3. Our model's relative error distribution pertinent to the nodal voltages in three three-phase DPSs with ZIP loads.

### C. Performance Against Bad Data

First, we tested the performance of our model in the case of bad training data. We employed the Huber penalty function (18) in (9). We replaced five groups of original training data with the new data where the voltage magnitude measurements were either very close to 0 or larger than 3.0, deviating badly from their normal ranges, i.e., 0.95 to 1.05 p.u. Then we redid the case studies on 22-bus, 85-bus, and 141-bus systems.

Fig. 4 shows the relative error distributions with and without using Huber penalty functions. Obviously, the Huber penalty function significantly improves our model's accuracy, ensuring its accuracy is robust against the bad data.

Second, we simulated the scenario where there were six bad data of either over 1.5 p.u. or below 0.5 p.u. in $\hat{v}_u$ and $\hat{v}_l$ on the 22-bus system. The motivation for this simulation was because one might wrongly take an outlier (but not severely deviating from a normal value) as $\hat{v}_u$ and $\hat{v}_l$ to represent very light or heavy loading conditions. In this simulation, we found that the average relative error without the Huber penalty function is $1.19 \times 10^{-2}$, and the error with the Huber penalty function is reduced to $6.1 \times 10^{-3}$. This result indicates that our linear model with the Huber penalty function employed in (9) would still have an acceptable accuracy against bad data even if these bad data were wrongly used in (7).

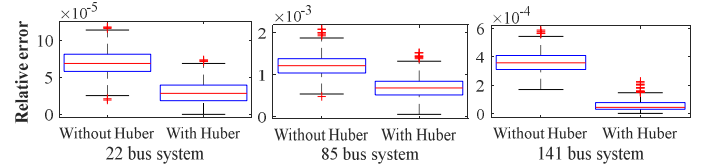

Fig. 4. Relative error distribution pertinent to the nodal voltages in the case of bad data.

### D. Application to Reactive Power Range Support Evaluation

We adopted the 5-bus system [35] and the 33-bus system from MATPOWER to show that our linear model facilitates a faster and relatively accurate evaluation of a DPS's reactive power support range.

Table IV lists the approximate support range evaluated with our linear model and the accurate evaluation using ACPF. The relatively small differences in $[\underline{q}_1, \overline{q}_1]$ and the shorter computational time demonstrate the efficacy of employing our linear model in this problem. In fact, even in the face of uncertainty, one can still employ our model in this reactive power range support evaluation problem simply by following the procedures in [37].

TABLE IV
REACTIVE POWER SUPPORT RANGE EVALUATION (TIME UNIT: SECONDS)

| $[\underline{q}_1, \overline{q}_1]$ (Mvar) /time[4] | Approximate Evaluation | Accurate Evaluation |
|---|---|---|
| 5 bus system | [-0.939, 9.341]/ $2.1 \times 10^{-3}$ | [-0.937, 9.346]/ $4.3 \times 10^{-3}$ |
| 33 bus system | [-9.827, 19.443]/ $2.6 \times 10^{-2}$ | [-9.836, 19.482]/ $6.1 \times 10^{-2}$ |

## V. CONCLUSION

This paper proposes a linear three-phase DPS power flow model applicable to both unbalanced and balanced DPSs with constant power or ZIP loads. We amalgamate data-driven techniques to improve the accuracy in linearizing a physics-based power flow model. Hence, our linear model is independent of some common assumptions in the literature and thus has a high accuracy generally. Case studies have demonstrated that ours is more accurate than other linear three-phase power flow models, with the average errors reduced by 2- to over 10-fold for different test systems.

Furthermore, this paper sheds light on several relevant issues. First, we show that using a Huber penalty function and forgetting factors reduce the adverse impact of bad data and further improve our model's accuracy. Second, we illustrate the necessity of employing the physics-based model on top of purely data-driven techniques. Besides, we demonstrate how to apply our model to a DPS's reactive power support range evaluation problem for a faster and relatively accurate solution.

---

[4] The codes were implemented on a PC with a 64-bit Intel Core i7 @2.9 GHz CPU and 16 GB of RAM.

Future works may involve considering the impact of possible interdependency in the historical operating data that are often assumedly independent in the literature. We will also continue to employ our linear model in other DPS optimization problems. These works will be reported in our future papers.